\documentclass{article}
\usepackage{arxiv}
\usepackage[utf8]{inputenc}
\usepackage[T1]{fontenc}
\usepackage{hyperref}
\usepackage{url}
\usepackage{booktabs}
\usepackage{amsfonts}
\usepackage{nicefrac}
\usepackage{microtype}
\usepackage{lipsum}
\usepackage{graphicx}
\usepackage{amsmath}
\usepackage{amssymb}
\usepackage{algorithm}
\usepackage{algorithmic}
\usepackage{natbib}

\title{Quantum State Preparation for Medical Data: Comprehensive Methods, Implementation Challenges, and Clinical Prospects}

\author{%
  \href{https://orcid.org/0000-0002-7268-0207}
  {
  Nikhil Kumar Rajput} \\
  Ramanujan College\\
  University of Delhi\\
  Delhi, India 110019 \\
  \texttt{n.rajput@ramanujan.du.ac.in} 
  \And
  \href{https://orcid.org/0009-0009-8983-1436}{
  Riya Bansal}
  \\
  Department of Computer Science\\
  University of Delhi\\
  Delhi, India 110007 \\
  \texttt{rbansal1@cs.du.ac.in} 
}

\date{}

\begin{document}
\maketitle

\begin{abstract}
Quantum computing holds transformative potential for medical applications, yet efficiently preparing quantum states from complex medical data remains a fundamental challenge. This survey provides a comprehensive examination of current approaches for encoding medical information into quantum systems, analyzing theoretical principles, algorithmic advancements, and practical limitations. It discusses tensor network decomposition, variational quantum algorithms, quantum machine learning techniques, and specialized error mitigation strategies for medical computing. The findings indicate that quantum advantages in medicine rely on leveraging inherent data structures such as spatial correlations in imaging, temporal patterns in physiological signals, and hierarchical biological organization. While current hardware restricts implementations to small-scale problems, emerging methods show potential for near-term use. The study provides a structured framework for assessing when quantum state preparation outperforms classical approaches in medicine, along with implementation guidelines and performance benchmarks.
\end{abstract}


\section{Introduction}

Quantum computing and medical informatics together are a promising but challenging area in computational science \citep{solenov2018potential}. Medical data is growing rapidly and comes in various forms. For instance, medical images contain millions of pixels with intricate spatial patterns that reflect anatomical structures. Genetic data follows patterns that reveal how genes work and relate to each other, while electronic health records (EHR) capture time-based trends related to disease development and treatment effects. These types of data are not only large but also highly structured and complex, making it difficult for classical computers to handle well \citep{tayefi2021challenges}. Thus, specialized approaches are required to handle them effectively. This opens the door for quantum computing. Quantum computers have the potential to handle large, structured data in ways that classical computers can't \citep{jeyaraman2024revolutionizing}. However, to unlock this potential, classical medical data must be encoded into a form that quantum computers can understand. This step is known as quantum state preparation. It involves translating detailed and structured medical data into quantum states, allowing for processing using quantum algorithms \citep{schuld2018supervised}.

A key difficulty in preparing medical data for quantum computing lies in the difference in scale between classical and quantum systems. Quantum computer uses qubits, and $n$-qubit quantum computer can represent $2^n$ possible states at once \citep{nielsen2010quantum}. This means that the amount of data they can process grows extremely fast. In contrast, classical medical data tends to grow in a slower and more structured way. This mismatch between the exponential capacity of quantum systems and the more linear nature of medical data creates both opportunities and challenges. Quantum computers could potentially compress and process large, complex medical datasets much more efficiently. But, building and using quantum systems that can take advantage of this is still very challenging with current technology.

Recent advances in quantum hardware, particularly in superconducting and trapped-ion systems have made near-term quantum applications increasingly viable \citep{arute2019quantum}. However, current devices still face limitations. They have a limited number of qubits available, are sensitive to noise, and can only operate for short periods before errors accumulate. These constraints are especially critical in medical applications, where high accuracy and reliability are essential. Despite these challenges, the field has witnessed remarkable progress. Researchers developed techniques to reduce errors \citep{fowler2012surface}, designing variational algorithms \citep{cerezo2021variational} that combine the strengths of classical and quantum computing, and exploring powerful new methods in quantum machine learning (QML) \citep{biamonte2017quantum}. Still, most applications in healthcare remain in the early stages and are primarily being studied in research labs, with real-world use cases yet to be fully realized.

This paper takes a deep look at how medical data can be prepared for use in quantum computers. We explore both the basic theory and the real-world challenges of turning complex medical data into quantum form. We review current methods, highlight areas where more research is needed, and offer practical advice for researchers and developers working in this field. Our discussion covers a range of healthcare applications, including drug discovery through molecular simulation, medical image analysis, genetic data processing, and personalized medicine.

\section{Mathematical Foundations}

Quantum state preparation for medical data relies on core mathematical principles that determine how efficiently and accurately data can be encoded. This section explores key concepts from information theory, computational complexity, and structural data properties to guide effective quantum encoding strategies.

\subsection{Information-Theoretic Framework}

From an information-theoretic perspective, quantum state preparation for medical data is fundamentally constrained by limits on how much classical information can be extracted from quantum systems. One key result is the Holevo bound \citep{holevo1973bounds}, which states that for a set of quantum states ${\rho_i, p_i}$, the maximum amount of classical information that can be accessed is bounded by:
\begin{equation}
\chi(\{p_i, \rho_i\}) = S\left(\sum_i p_i \rho_i\right) - \sum_i p_i S(\rho_i) \leq \log d
\end{equation}
Here, $S(\rho) = -\text{Tr}(\rho \log_2 \rho)$ is the von Neumann entropy, and $d$ is the dimension of the quantum system's Hilbert space. This bound implies that quantum states cannot encode unlimited classical data, which poses a critical challenge in medical applications where high fidelity and precise data representation are essential.

To address this, careful encoding strategies are needed, the ones that preserve the core medical information while taking advantage of quantum computational power. The von Neumann entropy not only measures information content but also helps assess how compressible a medical dataset is when transformed into quantum states. This is especially useful in fields like medical imaging, where structural patterns often exhibit spatial correlations. These correlations can be harnessed to optimize quantum encoding.

Additionally, in systems where medical data involves interacting parts, such as multiple physiological signals or paired genomic and phenotypic data, quantum mutual information becomes valuable. It measures how much information is shared between subsystems $A$ and $B$ and is given by:
\begin{align}
I(A:B)_\rho &= S(\rho_A) + S(\rho_B) - S(\rho_{AB}) \\
S(A|B)_\rho &= S(\rho_{AB}) - S(\rho_B)
\end{align}
These quantities help identify and exploit underlying correlations, guiding the development of encoding schemes that are not only efficient but also aligned with the intrinsic structure of the medical data.

\subsection{Complexity-Theoretic Considerations}

From a complexity-theoretic standpoint, quantum state preparation for medical data faces significant scalability challenges. In general, preparing an arbitrary quantum state on $n$ qubits requires resources that grow exponentially with $n$, following a complexity of $O(2^n)$. This exponential scaling presents a major obstacle for real-world medical applications, where datasets are often large and high-dimensional.

However, many types of medical data such as imaging, genomics, or physiological signals tend to have inherent structure and correlations. When the target quantum state exhibits limited entanglement or obeys certain structural constraints, the cost of preparing such a state can be reduced to polynomial scaling, or $O(\text{poly}(n))$. This suggests that not all medical datasets require prohibitively expensive quantum resources. Instead, structured medical data may be represented and encoded efficiently using quantum circuits with manageable complexity.

The key challenge lies in identifying which types of medical datasets possess such favorable structures. Recent advances in quantum complexity theory have shown that the entanglement properties of a quantum state are directly linked to the computational cost of preparing it \citep{eisert2010colloquium}. States with limited spatial correlations, hierarchical features, or low entanglement can often be constructed efficiently using classical preprocessing followed by quantum state preparation algorithms. This opens the door to practical applications in areas like MRI image encoding, genetic sequence analysis, or multi-modal patient data integration, provided these datasets exhibit the right kinds of structure for efficient quantum processing.

\subsection{Medical Data Structure Analysis}

Medical data exhibits several structural characteristics that make it well-suited for quantum encoding schemes. One prominent feature is spatial locality, especially in medical images, where adjacent pixels often contain similar intensity values due to the continuity of anatomical structures. This results in natural block structures that align well with tensor network representations, allowing for efficient decompositions that preserve clinically relevant features while significantly reducing computational complexity. Such spatial coherence is a foundational property of many imaging modalities, including MRI and CT, where anatomical regions exhibit smooth gradients and clearly defined boundaries.

In addition to spatial structure, hierarchical organization is a recurring theme in biological systems, from molecular assemblies to tissue-level architectures. This multi-scale nature mirrors the tree-like correlation patterns used in tree tensor networks, enabling efficient and compact quantum representations. By exploiting these hierarchical dependencies, quantum models can encode rich biological information in a scalable manner, which is particularly relevant in applications like multi-resolution medical imaging or layered biological analysis.

Furthermore, temporal dependencies are central to physiological monitoring applications. Signals such as electrocardiograms (ECG), electroencephalograms (EEG), and real-time patient monitoring data exhibit both short-term dynamics and long-term trends that are often nonlinear and complex. These sequential structures are well-matched to quantum encoding schemes designed for time-series data, where quantum states can represent temporal correlations compactly while preserving critical diagnostic features.

Beyond these structural properties, the quantum no-cloning theorem \citep{wootters1982single} introduces additional constraints specific to medical applications that must be carefully considered in algorithm design. Medical data privacy requirements align naturally with quantum information principles, as the no-cloning theorem provides inherent protection against unauthorized data replication. However, this constraint complicates classical machine learning (ML) workflows that rely on data replication for  training and validation. As a result, new QML frameworks must be designed to operate within this restriction, ensuring both data privacy and computational feasibility without compromising performance. These considerations highlight the necessity of domain-specific adaptations when applying quantum algorithms to real-world medical problems.

\section{Quantum State Preparation}

\begin{figure}
    \centering
    \includegraphics[width=0.8\linewidth]{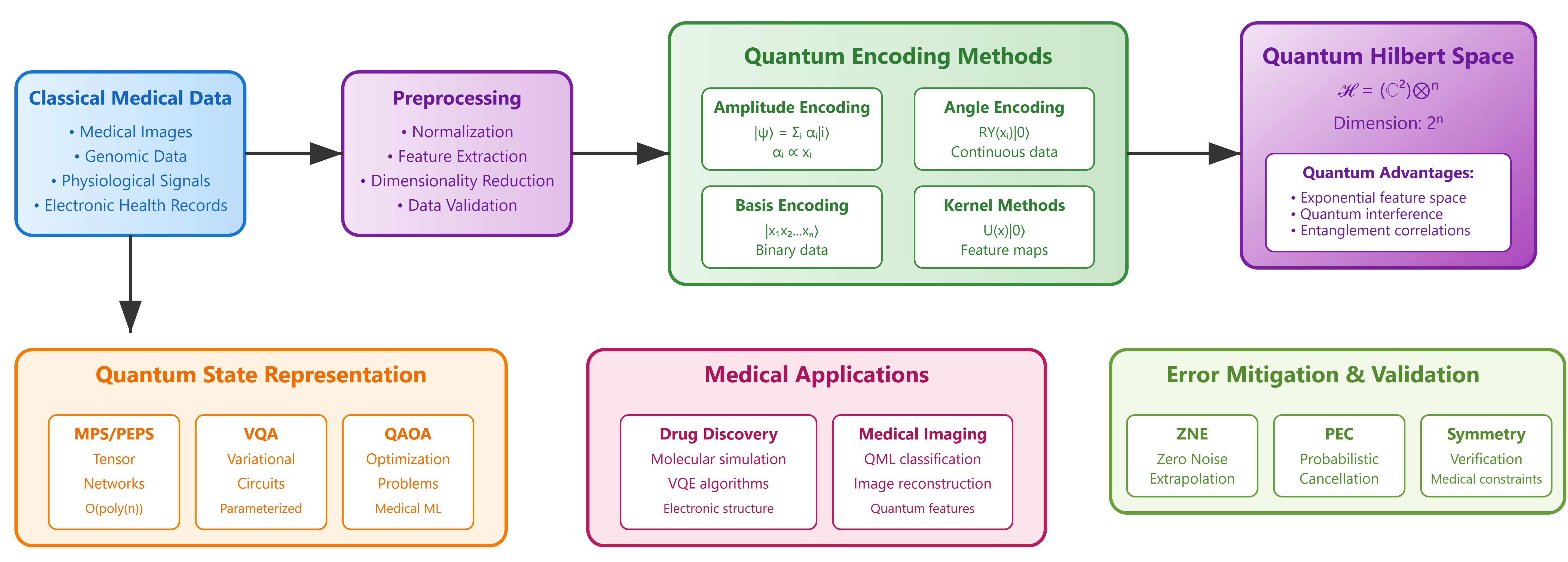}
    \caption{Quantum State Preparation for Medical Data}
    \label{fig:pipeline}
\end{figure}

The quantum state preparation for medical data follows a systematic workflow to transform classical healthcare information into quantum-encoded formats suitable for advanced computational analysis (Fig. \ref{fig:pipeline}). The process begins with comprehensive data acquisition from diverse medical sources including high-resolution imaging modalities (MRI, CT, X-ray), electrophysiological signals (ECG, EEG), genomic sequences, and structured EHR. This raw data undergoes rigorous preprocessing involving specialized normalization techniques tailored for medical data types, advanced feature extraction, and dimensionality reduction while preserving clinically relevant patterns. 

The quantum encoding phase employs multiple parallel strategies to optimally represent medical data in quantum systems. Amplitude encoding efficiently captures probability distributions in diagnostic data, while angle encoding better represents periodic physiological signals. Basis encoding handles categorical clinical variables, and kernel encoding creates enhanced feature spaces for machine learning applications. These methods collectively map the medical data into an exponentially large quantum Hilbert space $\mathcal{H} = (\mathbb{C}^2)^{\otimes n}$, where the $2^n$-dimensional state space enables simultaneous processing of complex medical patterns through quantum superposition and entanglement. This fundamental quantum advantage allows for unprecedented analysis of multidimensional medical relationships that remain computationally intractable for classical systems.  

Within the quantum processing core, specialized algorithms perform medical-specific computations. Tensor network methods efficiently represent high-dimensional medical data structures, while variational quantum algorithms (VQA) optimize parameterized circuits for specific clinical tasks. The Quantum Approximate Optimization Algorithm (QAOA) solves complex treatment planning problems, and QML models enhance diagnostic pattern recognition. For drug discovery applications, variational quantum eigensolvers simulate molecular interactions at unprecedented scales. Further, it incorporates robust error mitigation through advanced techniques like zero-noise extrapolation and probabilistic error cancellation to ensure reliable outputs despite current hardware limitations.

\section{Quantum Data Encoding for Medical Applications} \label{quantum_feature_map}

Quantum feature maps form the foundation of QML by systematically encoding classical medical data into quantum states. The choice of encoding strategy is crucial, as it affects not only the computational efficiency but also the ability to leverage quantum advantages in downstream applications. Different types of medical data ranging from discrete genetic markers to continuous physiological signals require distinct encoding techniques tailored to their structure and semantics.

Basis encoding is the most straightforward method, where classical binary data is directly mapped to computational basis states of qubits. This encoding is particularly suitable for discrete medical data such as genetic variants, diagnostic codes, or categorical patient information. The encoding maps classical data $x = x_1 x_2 \ldots x_n$ to quantum states $|x\rangle = |x_1 x_2 \ldots x_n\rangle$, preserving the classical information structure while enabling quantum processing.

Amplitude encoding offers the potential for exponential compression by encoding classical data in the amplitudes of quantum states. For normalized classical data $x$, the amplitude encoding creates quantum states of the form:
\begin{equation}
|\psi(x)\rangle = \frac{1}{\|x\|} \sum_{i=1}^{2^n} x_i |i\rangle
\end{equation}
This encoding can represent exponentially large classical datasets using linearly growing numbers of qubits, though the state preparation overhead often negates this advantage in practice.

Angle encoding is well-suited for continuous-valued medical data, such as laboratory results, heart rate, or blood pressure readings. It encodes each feature $x_i$ as a rotation angle, producing single-qubit states of the form:
\begin{equation}
|\psi(x)\rangle = \bigotimes_{i=1}^n \left(\cos(x_i)|0\rangle + \sin(x_i)|1\rangle\right)
\end{equation}
This method is relatively hardware-efficient and interpretable, making it ideal for applications involving sensor data or time-series physiological signals.

Beyond these basic encodings, quantum kernel methods offer a more expressive framework by employing parameterized quantum circuits to construct complex, high-dimensional feature spaces. These methods define quantum feature maps as 
$\phi(x) = U_{\text{feature}}(x)|0\rangle^{\otimes n}$, where the unitary $U_{\text{feature}}(x)$ is designed to transform classical data into a quantum state whose inner products define a quantum kernel. Such feature spaces may be classically intractable to simulate, potentially unlocking quantum advantages in classification or regression tasks within precision medicine and biomedical diagnostics.

\section{Tensor Network Methods} \label{tensor}

\subsection{Matrix Product States for Medical Imaging}

\begin{figure}
    \centering
    \includegraphics[width=0.8\linewidth]{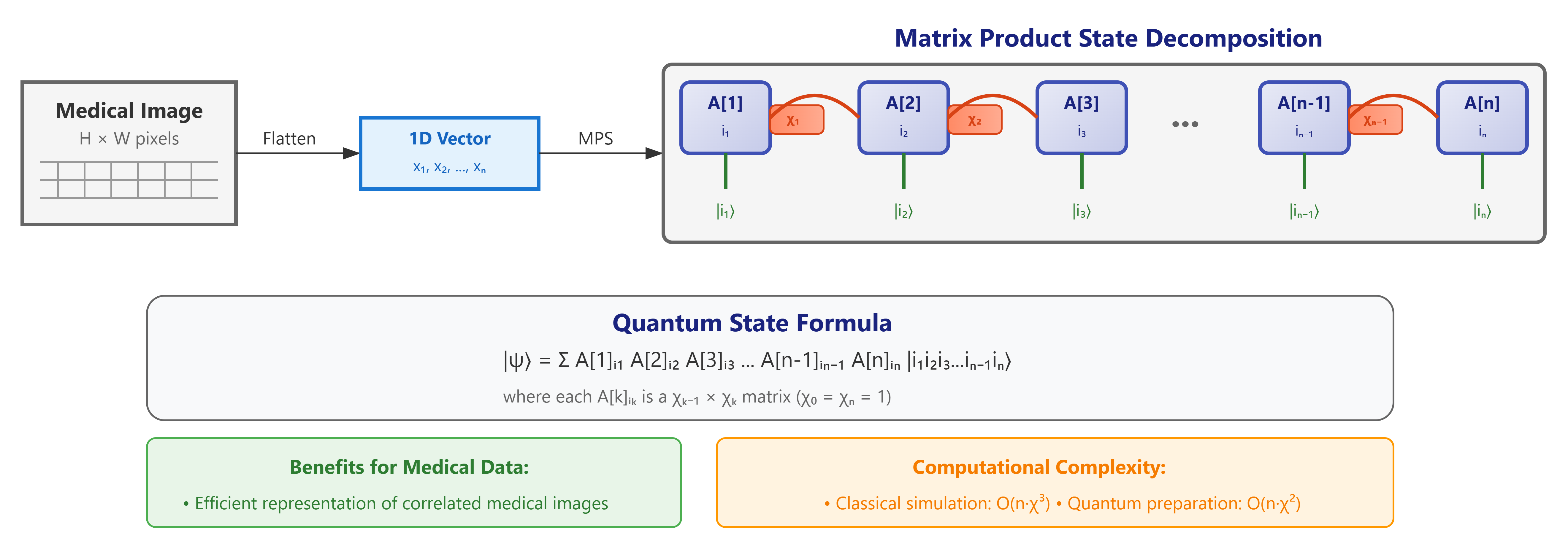}
    \caption{Matrix Product State for Medical Image Data}
    \label{fig:mps}
\end{figure}

Matrix Product States (MPS) provide systematic frameworks for encoding structured medical data into quantum representations \citep{orus2014practical}. This frameworks is illustrated in Fig. \ref{fig:mps}. According to this figure, when medical image converted to 1D vector, MPS is applied. This results into MPS decomposition which expresses a quantum state as a chain of connected tensors, each describing local correlations while maintaining global coherence through entanglement bonds. For a quantum state representing medical data, the MPS decomposition takes the form:
\begin{equation}
|\psi\rangle = \sum_{i_1,\ldots,i_n} A^{[1]}_{i_1} A^{[2]}_{i_2} \cdots A^{[n]}_{i_n} |i_1 i_2 \ldots i_n\rangle
\end{equation}
where each tensor $A^{[k]}_{i_k}$ has dimensions $\chi_{k-1} \times \chi_k$, and the bond dimensions $\chi_k$ control the expressiveness of the representation.

For medical imaging applications, MPS methods require careful consideration of mapping 2D image data to 1D tensor chains. The choice of ordering scheme significantly affects the efficiency of the MPS representation and determines which spatial correlations are preserved during the decomposition process. Row-major ordering preserves horizontal correlations in medical images, making it suitable for applications where horizontal anatomical features are important. Column-major ordering preserves vertical correlations, which may be more appropriate for images with predominantly vertical structures. Space-filling curves attempt to preserve both horizontal and vertical locality by following paths that minimize the distance between adjacent pixels in the 1D mapping. Recursive bisection hierarchically partitions image regions, creating balanced decompositions that preserve multi-scale structure.

The approximation quality of MPS representations depends fundamentally on the entanglement structure of the target quantum state. For medical data with limited correlation length $\xi$, theoretical bounds exist on the approximation error achieved by MPS decompositions with finite bond dimension. The relationship between approximation error and bond dimension follows:
\begin{equation}
\|\rho_{\text{exact}} - \rho_{\text{MPS}}\|_1 \leq 2\sqrt{2\epsilon_{\text{SVD}}}
\end{equation}
where $\epsilon_{\text{SVD}}$ represents the truncation error in the singular value decomposition used to construct the MPS representation. The following process of MPS construction for medical image data is outlined in Algorithm \ref{alg:MPS}.

Recent work has explored specialized MPS variants for medical data that exploit specific structural properties. Symmetric MPS can exploit bilateral symmetry common in anatomical structures, reducing the number of parameters required for accurate representation. Infinite MPS approaches handle streaming medical data with translational invariance properties, enabling real-time processing of continuous physiological monitoring data.

\begin{algorithm}
\scriptsize
\caption{MPS Construction for Medical Image Data}
\begin{algorithmic}[1]
\REQUIRE Medical image data $I \in \mathbb{R}^{H \times W}$, target approximation error $\epsilon$
\ENSURE MPS tensors $\{A^{[k]}\}$ and quantum circuit
\STATE Choose ordering scheme based on correlation analysis
\STATE Reshape image to vector: $v = \text{flatten}(I)$
\STATE Normalize vector: $v \leftarrow v/\|v\|$
\STATE Initialize tensor decomposition
\FOR{$k = 1$ to $n-1$}
    \STATE Reshape current data to matrix form
    \STATE Perform SVD: $M = U\Sigma V^\dagger$
    \STATE Determine truncation rank from target error $\epsilon$
    \STATE Store left singular vectors as $A^{[k]}$
    \STATE Update remaining data: $M_{\text{next}} = \Sigma V^\dagger$
\ENDFOR
\STATE Store final tensor as $A^{[n]}$
\STATE Compile MPS to quantum circuit
\RETURN MPS tensor sequence and quantum circuit
\end{algorithmic}
\label{alg:MPS}
\end{algorithm}

\subsection{Tree Tensor Networks for Hierarchical Medical Data}

Tree Tensor Networks (TTN) provide natural frameworks for representing medical data with inherent hierarchical structure \citep{shi2006classical}. Biological systems exhibit hierarchical organization from molecular to cellular to tissue levels, creating natural alignments with TTN representations that can capture multi-scale correlations efficiently. Medical classification systems and anatomical taxonomies also possess tree-like structures that align naturally with TTN decompositions.

The TTN representation organizes tensors in a tree structure where each node represents a local tensor describing correlations among its connected edges. The global quantum state is obtained by contracting all tensors according to the tree topology:
\begin{equation}
|\psi\rangle = \sum_{\{i\}} T^{(\text{tree})}_{i_1,\ldots,i_N} |i_1, \ldots, i_N\rangle
\end{equation}
where the tensor $T^{(\text{tree})}$ is defined by contractions over the tree graph structure.

The tree structure enables TTNs to efficiently model long-range correlations, outperforming MPS representations that decay exponentially with distance. Such capabilities are especially valuable in medical contexts like whole-body imaging or multi-organ physiological monitoring, where global dependencies across distant regions are diagnostically relevant.

In terms of computational resources, the TTN construction scales as $O(N\chi^3)$, where $N$ is the number of data points and $\chi$ is the maximum bond dimension. This polynomial complexity offers a favorable trade-off between efficiency and expressivity, making TTNs practical for moderately sized medical datasets while still preserving intricate correlation structures critical for downstream quantum processing.

\subsection{Projected Entangled Pair States for 2D Medical Images}

For inherently 2D medical images, Projected Entangled Pair States (PEPS) \citep{verstraete2008matrix} provide more natural representations than MPS by preserving the native spatial structure of the data. PEPS represent quantum states using a network of tensors arranged on a 2D lattice, where each tensor corresponds to a pixel or image region:
\begin{equation}
|\psi\rangle = \sum_{\{i\}} \text{Tr}\left[\prod_{\text{vertices}} A^{[v]}_{i_v}\right] |i_1, \ldots, i_N\rangle
\end{equation}

PEPS representations preserve 2D spatial structure while maintaining polynomial scaling in bond dimension, making them particularly suitable for medical imaging applications where spatial relationships are crucial for preserving diagnostic information. However, PEPS contraction is generally \#P-complete, requiring approximation algorithms for practical implementation. Recent advances in approximate contraction algorithms have made PEPS more practical for medical applications, though computational overhead remains significant compared to MPS approaches.

\subsection{Matrix Product Operators for Medical Image Processing}

Matrix Product Operators (MPO) extend the MPS framework to represent operations on quantum states rather than states themselves \citep{crosswhite2008finite}. This extension is particularly relevant for medical image processing operations such as filtering, edge detection, and feature extraction, which can be represented as MPO applied to image states encoded as MPS.

The MPO representation expresses an operator as a product of local operators connected by bond indices:
\begin{equation}
O = \sum_{i,j,k,l} W^{[1]}_{i,j} W^{[2]}_{k,l} \cdots |i, k, \ldots\rangle\langle j, l, \ldots |
\end{equation}

Common medical image processing operations can be decomposed into MPO form with varying degrees of efficiency. Convolution operations enable spatial filtering for noise reduction and edge detection with bond dimensions that depend on filter size and structure. Fourier transforms provide frequency domain analysis for medical image enhancement, though they typically require large bond dimensions due to their global nature. Wavelet transforms offer multi-scale analysis for feature extraction with more favorable bond dimension scaling. Morphological operations support shape-based image processing for segmentation with bond dimensions determined by structuring element complexity.

The bond dimension requirements for representing these operations determine their feasibility for quantum implementation on near-term devices. Operations with small, local support typically admit efficient MPO representations, while global operations may require prohibitively large bond dimensions for practical quantum implementation.

\section{Variational Quantum Algorithms} \label{vqa}

\subsection{Variational Quantum Eigensolver for Molecular Medicine}

The Variational Quantum Eigensolver (VQE) \citep{peruzzo2014variational} represents one of the most promising near-term quantum algorithms for medical applications, particularly in molecular simulation for drug discovery. The VQE approach constructs parameterized quantum circuits to approximate ground state solutions of molecular Hamiltonians, providing a bridge between current noisy quantum devices and scientifically relevant molecular systems.

The variational approach constructs parameterized quantum circuits to approximate target quantum states through classical optimization of circuit parameters. The general VQE framework involves preparing trial states of the form:
\begin{align}
|\psi(\theta)\rangle &= U(\theta)|0\rangle^{\otimes n} \\
L(\theta) &= \langle\psi(\theta)|H|\psi(\theta)\rangle + \lambda R(\theta)
\end{align}
where $U(\theta)$ represents a parameterized quantum circuit and $R(\theta)$ provides regularization to prevent overfitting and improve convergence properties.

For molecules relevant to drug design, the Hamiltonian is expressed in second quantization as:
\begin{equation}
H = \sum_{pq} h_{pq} a^\dagger_p a_q + \frac{1}{2}\sum_{pqrs} h_{pqrs} a^\dagger_p a^\dagger_q a_r a_s
\end{equation}
where $h_{pq}$ and $h_{pqrs}$ are one- and two-electron integrals, and the operators $a^\dagger$, $a$ correspond to fermionic creation and annihilation.

Since quantum hardware operates on qubits rather than fermions, fermion-to-qubit transformations are essential. The Jordan-Wigner transformation maps fermionic operators to Pauli operators as:
\begin{align}
a_j = \frac{1}{2}(X_j + iY_j) \prod_{k=0}^{j-1} Z_k,  \quad \quad \quad
a^\dagger_j = \frac{1}{2}(X_j - iY_j) \prod_{k=0}^{j-1} Z_k
\end{align}
This method introduces long strings of 
$Z$ operators, potentially increasing circuit depth. To mitigate this, the Bravyi-Kitaev transformation \citep{bravyi2002fermionic} offers an alternative mapping that reduces the number of Pauli terms in the mapped Hamiltonian:
\begin{equation}
a_j = \frac{1}{2}(X_j + iY_j) \prod_{k \in P(j)} Z_k
\end{equation}
where $P(j)$ denotes the parity set defined by binary tree structure. This representation often results in fewer Pauli terms, reducing circuit complexity and improving execution efficiency on Noisy Intermediate-Scale Quantum (NISQ) hardware.

\subsection{Barren Plateau Mitigation for Medical Applications}

The barren plateau phenomenon poses a fundamental challenge for VQA, particularly in the context of quantum state preparation \citep{mcclean2018barren}. As the number of qubits increases, the gradient of the cost function with respect to the circuit parameters tends to vanish exponentially, leading to flat optimization landscapes that hinder convergence. For many generic parameterized quantum circuits, the gradient variance scales as:
\begin{equation}
\text{Var}[\nabla L] \leq \frac{C}{4^n}
\end{equation}
where $C$ is a constant and $n$ denotes the number of qubits, making optimization exponentially difficult as system size grows.

This issue is particularly acute in medical applications of quantum computing, where medical tasks like studying large molecules, high-quality medical images, or full patient records usually need big quantum systems. Research shows that as the system size gets bigger, the chances of avoiding this problem get much smaller, making it hard to use quantum computing effectively in real-world medical cases.

To address these challenges, several mitigation strategies have been proposed. One approach involves using problem-specific ansätze that encode known physical or biological structure from the medical domain, thus constraining the parameter space to regions less prone to barren plateaus. Initialization strategies that begin optimization from classically-informed starting points can improve convergence by providing good initial approximations to optimal parameters. Hierarchical approaches that build complex states from simpler components can circumvent some scaling challenges by decomposing large optimization problems into smaller, more manageable subproblems. Further, parameter correlation analysis has revealed that many medical applications exhibit structured parameter landscapes where correlations between parameters can guide optimization strategies. These correlations often reflect underlying physical or biological constraints that can be exploited to improve optimization efficiency and avoid barren plateau regions.

\subsection{Hardware-Efficient Ansätze}

Hardware-efficient ansätze are a class of parameterized quantum circuits specifically designed to align with the physical constraints of current quantum hardware, such as limited qubit connectivity and native gate sets \citep{kandala2017hardware}. These ansätze aim to balance expressiveness with implementability, creating circuits that can represent complex quantum states while remaining executable on near-term quantum hardware. The general form of hardware-efficient ansätze involves alternating layers of single-qubit rotations and two-qubit entangling gates. This structure can be expressed as:
\begin{equation}
U_{\text{HEA}}(\theta) = \prod_{l=1}^L \left[\prod_{i=1}^n R_Y(\theta_{i,l}) \prod_{\langle j,k \rangle} \text{CZ}_{j,k}\right]
\end{equation}
where entangling gates are applied according to the device’s native connectivity graph, and the circuit depth $L$ must balance expressiveness with resilience to noise.

In medical quantum computing, designing hardware-efficient ansätze introduces challenges as well as opportunities. Medical datasets often exhibit rich and domain-specific correlation structures. These correlations may not map naturally onto the qubit connectivity graphs of current quantum processors. As a result, implementing a meaningful ansatz for medical tasks may require SWAP operations to emulate non-local interactions, or the development of customized ansätze that better reflect both the data structure and hardware constraints. Optimizing this alignment is critical since overly shallow circuits may fail to capture essential features, while deeper circuits risk degradation from quantum noise. Therefore, effective use of hardware-efficient ansätze in medicine demands careful co-design of the quantum model, the encoding strategy, and the target hardware platform.

\subsection{Problem-Specific Ansätze for Medical Applications}

Problem-specific ansätze are tailored quantum circuit designs that incorporate domain knowledge from medical and biological sciences to enhance optimization efficiency and avoid challenges like barren plateaus \citep{grimsley2019adaptive}. By embedding known structure into the circuit architecture, these ansätze restrict the search space to physically meaningful configurations, making training more efficient and interpretable. For instance, symmetry-preserving ansätze take advantage of spatial or temporal symmetries, commonly found in medical data. These symmetries reduce the number of trainable parameters and help distinguish true computational signals from artifacts or errors.

In molecular medicine and drug discovery, chemistry-inspired ansätze utilize the known structure of molecular orbitals to represent quantum states relevant to chemical interactions. One notable example is the Unitary Coupled Cluster (UCC) ansatz, which uses chemical intuition to generate electron excitations from a reference state:
\begin{equation}
|\psi_{\text{UCC}}\rangle = e^{T - T^\dagger} |\phi_0\rangle
\end{equation}
Here, $T$ encodes excitations from a reference state $|\phi_0\rangle$ based on chemical intuition about electron correlation patterns. 

In imaging applications, image-aware ansätze are designed to respect spatial locality, mirroring how diagnostic features like tumors or lesions are concentrated in specific anatomical regions. These circuits are structured to capture local correlations, reducing noise and improving model focus on clinically relevant areas. For time-series data, such as ECG or EEG signals, temporal ansätze are employed. These circuits are constructed to capture both short-term events (like arrhythmias) and long-term trends (such as circadian rhythms). Often, they include elements analogous to classical recurrent networks, enabling the retention and processing of sequential dependencies while maintaining quantum coherence. In all these cases, integrating domain-specific insights into the ansatz structure ensures that the quantum model aligns more closely with the medical problem at hand, improving training efficiency, interpretability, and real-world relevance.

\section{Quantum Machine Learning for Medical Data}

\subsection{Variational Quantum Classifiers for Medical Diagnosis}

\begin{figure}
    \centering
    \includegraphics[width=0.8\linewidth]{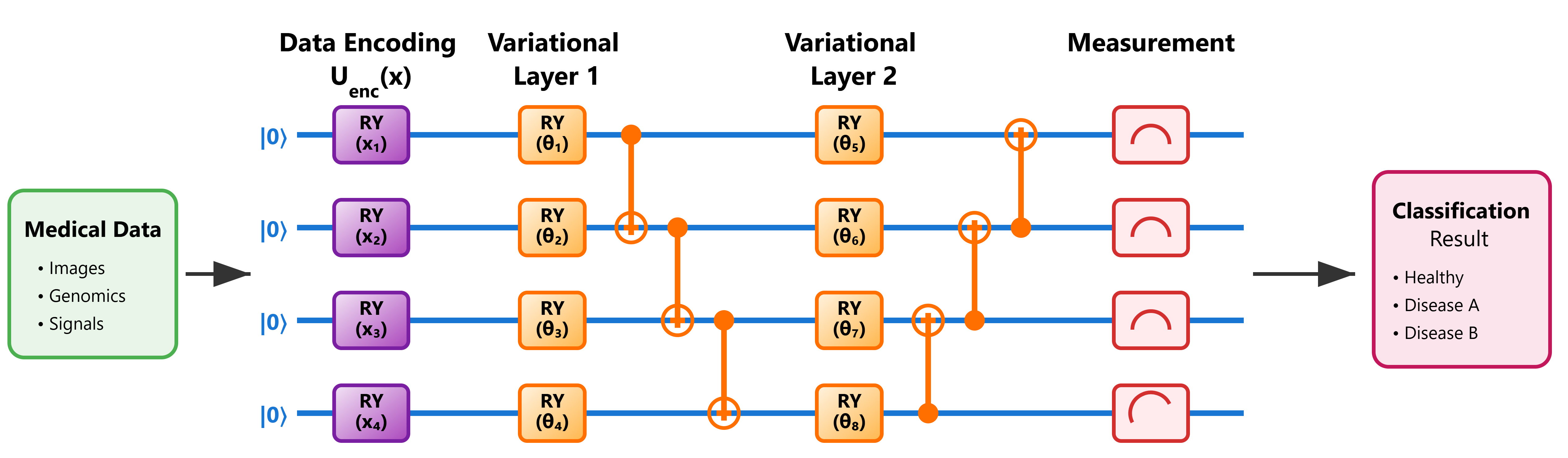}
    \caption{Variational Quantum Circuit for Medical Data Classification}
    \label{fig:vqc}
\end{figure}

Variational Quantum Classifiers (VQCs) represent a promising approach in QML by integrating classical data encoding with parameterized quantum circuits, offering potential advantages for medical diagnosis tasks \citep{mitarai2018quantum}. The VQC framework distinctly separates the encoding of classical medical data from the variational processing required for classification, enabling flexible adaptation to different medical data types and diagnostic tasks. As illustrated in Fig. \ref{fig:vqc}, the architecture of a VQC typically consists of an encoding circuit $U_{enc}(x)$, which maps classical input $x$ into a quantum state, followed by a variational circuit $U_{var}(\theta)$ parameterized by tunable variables $\theta$. The overall quantum state is thus represented as:
\begin{equation}
|\psi(x,\theta)\rangle = U_{\text{var}}(\theta) U_{\text{enc}}(x) |0\rangle^{\otimes n}
\end{equation}

Classification is performed by measuring appropriate observables on the final quantum state, with the outcomes interpreted probabilistically to yield predicted labels. Training the VQC involves optimizing the parameters $\theta$ to minimize a loss function, typically derived from cross-entropy or hinge loss, using quantum-compatible gradient-based methods. One of the key techniques enabling efficient training on quantum hardware is the parameter shift rule, which allows computation of gradients without requiring full quantum state tomography \citep{schuld2019evaluating}: 
\begin{equation}
\frac{\partial}{\partial \theta_i} \langle O \rangle = \frac{1}{2}[\langle O \rangle_{\theta_i + \pi/2} - \langle O \rangle_{\theta_i - \pi/2}]
\end{equation}
This rule facilitates direct hardware-based training by evaluating expectation values at shifted parameter points. However, practical deployment of VQCs in healthcare is currently limited by the shallow circuit depths supported by NISQ devices. Many medical classification problems demand complex feature transformations that surpass the expressive capacity of shallow circuits. Furthermore, quantum measurement noise and decoherence can significantly degrade the precision of classification outcomes, posing challenges in applications where high diagnostic accuracy is critical. Despite these limitations, VQCs remain a compelling direction for future research, especially as quantum hardware continues to advance.

\subsection{Quantum Generative Models for Medical Data Synthesis}

Quantum generative models offer exciting new ways to create synthetic medical data. These models use quantum properties like superposition and entanglement to potentially learn complex data patterns that classical methods may struggle with. This is especially useful in healthcare, where generating synthetic data can help with training ML models, protecting patient privacy, and increasing the size of datasets through augmentation.

One popular approach is the Quantum Generative Adversarial Network (qGAN) \citep{zoufal2019quantum}, which is the quantum version of the well-known GAN used in classical ML. It includes two main parts: a quantum generator, which creates fake medical data from random quantum inputs, and a quantum discriminator, which tries to tell the difference between real and fake data. These two components compete with each other in a training loop, aiming to improve the quality of the synthetic data. The goal is to find a balance between the generator and discriminator by minimizing and maximizing a shared loss function:
\begin{equation}
\min_{\theta_G} \max_{\theta_D} \mathbb{E}_{|\psi\rangle \sim p_{\text{data}}}[\log D(|\psi\rangle)] + \mathbb{E}_{|z\rangle \sim p_z}[\log(1 - D(G(|z\rangle)))]
\end{equation}
However, training qGANs is not easy. Even classical GANs are known to be difficult to train, and the quantum version faces additional challenges. These include noise in quantum measurements, limited expressiveness of small quantum circuits, and difficulties in calculating smooth gradients due to the discrete nature of quantum outputs.

Another promising option is the Quantum Born Machine \citep{liu2018differentiable}, which takes a simpler approach. Instead of using a discriminator, it directly models the probability of generating each data point using a quantum circuit. The output distribution is determined by the square of the amplitudes of the quantum state:
\begin{equation}
p(x) = |\langle x|\psi(\theta)\rangle|^2
\end{equation}
This makes Born machines more stable to train compared to qGANs, though they still rely on having enough qubits and low-noise hardware to work well.

Currently, these quantum generative models are limited by the capabilities of today’s quantum computers. Since medical datasets are large and complex, modeling such data would require deep and expressive quantum circuits. As a result, these models are mainly tested on small or simplified datasets. Still, as quantum hardware improves, these models could play a major role in generating realistic medical data in the future.

\subsection{Quantum Autoencoders for Medical Data Compression}

Quantum autoencoders are QML models designed to reduce the size of quantum representations of medical data while preserving key information \citep{romero2017quantum}. These models are useful in medical applications where data can be large and complex and need to be compressed for efficient processing on quantum hardware.
The architecture of a quantum autoencoder includes two components:
\begin{itemize}
    \item The encoder, which transforms the input quantum state (representing medical data) into a compressed, lower-dimensional state.

        \text{Encoder:} \quad E : $\mathcal{H}_{\text{data}} \rightarrow \mathcal{H}_{\text{latent}}$

    \item The decoder, which reconstructs the original input from this compressed representation.

        \text{Decoder:} \quad D : $\mathcal{H}_{\text{latent}} \rightarrow \mathcal{H}_{\text{data}}$
    
\end{itemize}
The goal is to find an optimal way to compress the data so that the output of the decoder closely matches the original input. This is achieved by minimizing a loss function that includes both the reconstruction error and a regularization term to ensure the learned representations are meaningful and generalizable:
\begin{equation}
L_{\text{autoencoder}} = \|D(E(|\psi_{\text{data}}\rangle)) - |\psi_{\text{data}}\rangle\|^2 + \lambda R(E(|\psi_{\text{data}}\rangle))
\end{equation}

To evaluate how well the reconstructed state matches the original, quantum autoencoders use a method called the swap test. This test measures the fidelity between two quantum states (how similar they are) using a quantum circuit with an ancilla qubit:
\begin{equation}
F = |\langle\psi_{\text{original}}|\psi_{\text{reconstructed}}\rangle|^2 = 2\Pr[\text{ancilla} = |0\rangle] - 1
\end{equation}
The swap test offers a major advantage: it allows for accurate fidelity estimation without the need for full quantum state tomography, which would otherwise require a large number of measurements. This results in quadratic savings in the number of measurements needed, making quantum autoencoders more practical on current quantum devices that have limited measurement and noise-handling capabilities.

\section{Specialized Preparation Methods} \label{qaoa}

\subsection{Quantum State Tomography for Medical Data Validation}

Quantum state tomography plays a vital role in validating the accuracy of quantum states used in medical applications. It allows researchers to verify whether quantum algorithms have successfully encoded classical medical data into accurate quantum representations. This process involves reconstructing the full quantum state, typically represented by a density matrix, using statistical data obtained from repeated measurements of various quantum observables. However, full quantum state tomography is computationally expensive. For a system with $n$ qubits, the number of required measurements grows exponentially, on the order of $O(4^n)$. This exponential scaling makes full tomography impractical for large quantum systems, especially when dealing with high-dimensional medical data, which may involve multiple variables and large datasets. As a result, performing traditional tomography on quantum states that encode medical information quickly becomes infeasible with current hardware.

To address this, compressed sensing tomography offers a more efficient alternative, particularly suitable for medical quantum states that often possess special structure (e.g., sparsity or low-rank behavior) \citep{gross2010quantum}. This method recasts tomography as an optimization problem, seeking the most likely quantum state that matches the measurement data, while assuming the state has a low-rank structure:
\begin{equation}
\min_\rho \text{Tr}(\rho) \quad \text{subject to} \quad \|\mathcal{A}(\rho) - b\|_2 \leq \epsilon
\end{equation}
Here, $\rho$ is the quantum state (density matrix) to be reconstructed, $\mathcal{A}$ is a linear map representing the measurement process, and $b$ is the vector of observed measurement results. The constraint ensures the reconstructed state is consistent with the data within a small error margin $\epsilon$.

In the context of medical quantum computing, prior domain knowledge can be leveraged to further enhance efficiency. For example, many medical datasets exhibit structured correlations, such as spatial locality in medical images, gene-gene interactions in genomics, or time dependencies in patient monitoring. These structures can guide the selection of measurement bases and reconstruction techniques, helping to reduce the number of required measurements and improve reconstruction accuracy.

\subsection{Adiabatic Quantum State Preparation}

Adiabatic quantum state preparation provides an alternative to VQA by slowly evolving a quantum system from a simple, known state to a more complex one that encodes medical data \citep{albash2018adiabatic}. This approach is based on the adiabatic theorem in quantum mechanics, which states that if a system starts in the ground state of an initial Hamiltonian and the system evolves slowly enough, it will remain in the ground state of the Hamiltonian throughout the evolution.

In this method, a time-dependent Hamiltonian is defined that gradually changes from an initial form $H_0$ to a final, problem-specific Hamiltonian $H_1$:
\begin{equation}
H(t) = (1-s(t))H_0 + s(t)H_1
\end{equation}
Here, $s(t)$ is a smooth scheduling function that increases from 0 to 1 as time progresses. The Hamiltonian $H_0$ has a simple ground state that is easy to prepare on a quantum computer, while $H_1$ encodes the structure of the medical problem.

To successfully follow the ground state during this evolution, the process must be slow enough, based on the minimum energy gap $\Delta(t)$ between the ground and excited states:
\begin{equation}
T \gg \frac{\max_t |\langle 1(t)|\partial_t H(t)|0(t)\rangle|}{\Delta(t)^2}
\end{equation}
The smaller the energy gap $\Delta(t)$ , the longer the total runtime $T$ must be to ensure the system stays in the correct state. This can pose a challenge, since long runtimes may exceed the quantum coherence time, the period over which quantum systems can maintain their state without error, especially on today’s noisy quantum devices.

Despite this limitation, adiabatic approaches are appealing in medical settings where quantum states must encode complex correlations or interdependent variables. These are often difficult to represent efficiently using variational methods, which rely on predefined circuit templates (ansätze).

\subsection{Quantum Approximate Optimization Algorithm}

The Quantum Approximate Optimization Algorithm (QAOA) \citep{farhi2014quantum} offers a systematic method for preparing quantum states tailored to solving optimization problems frequently encountered in medical domains. QAOA constructs parameterized quantum states through alternating application of problem-specific and mixing Hamiltonians:
\begin{equation}
|\psi(\gamma, \beta)\rangle = \prod_{p=1}^P e^{-i\beta_p H_M} e^{-i\gamma_p H_C} |+\rangle^{\otimes n}
\end{equation}
where $H_C$ encodes the cost function of the optimization problem and $H_M$ provides mixing to explore the solution space.

Medical optimization applications that benefit from QAOA approaches include treatment planning for optimizing radiation therapy dose distributions, drug design for molecular optimization problems, resource allocation for hospital scheduling and logistics, and clinical trial design for patient stratification and protocol optimization. These applications often involve combinatorial optimization problems that are well-suited to the QAOA framework.

The performance of QAOA depends critically on the choice of circuit depth $P$ and the optimization of angle parameters $\gamma$ and $\beta$. Deeper circuits can represent more complex solution structures but require longer coherence times and introduce additional noise. The optimization landscape for QAOA parameters can exhibit barren plateaus similar to other VQA, though problem-specific structure may provide advantages for certain medical optimization problems.

\subsection{Machine Learning-Optimized State Preparation}

Recent developments have introduced ML techniques to improve quantum state preparation, framing the design of quantum circuits as a learning problem. In this approach, models such as neural networks and reinforcement learning (RL) agents are trained to discover efficient quantum circuits tailored to specific medical applications \citep{fosel2018reinforcement}.

In RL-based methods, the process of building a quantum circuit is viewed as a sequential decision-making task. An RL agent interacts with an environment where each state represents the current circuit configuration, and actions correspond to adding quantum gates or adjusting gate parameters. The agent learns a policy that maximizes a reward function designed to balance accuracy and efficiency:
\begin{equation}
R = \alpha \cdot \text{fidelity} - \beta \cdot \text{circuit depth} - \gamma \cdot \text{gate count}
\end{equation}
This formulation encourages the agent to prepare quantum states with high fidelity while keeping the circuits compact and practical to implement on real hardware.

Meanwhile, neural network approaches focus on learning the mapping between patterns in classical medical data, and the optimal parameters or structures of quantum circuits. These models can automatically adapt quantum state preparation strategies to different data types and might discover unconventional circuit designs that outperform standard, hand-crafted ansätze in specific medical contexts.

However, the training requirements for ML-optimized approaches can be substantial, requiring large numbers of quantum circuit evaluations that may be impractical on near-term quantum devices. Hybrid approaches that combine classical optimization with quantum evaluation may provide more practical pathways for implementing learned state preparation strategies.

\section{Error Mitigation for Medical Quantum Computing} \label{error}

\begin{figure}
    \centering
    \includegraphics[width=0.8\linewidth]{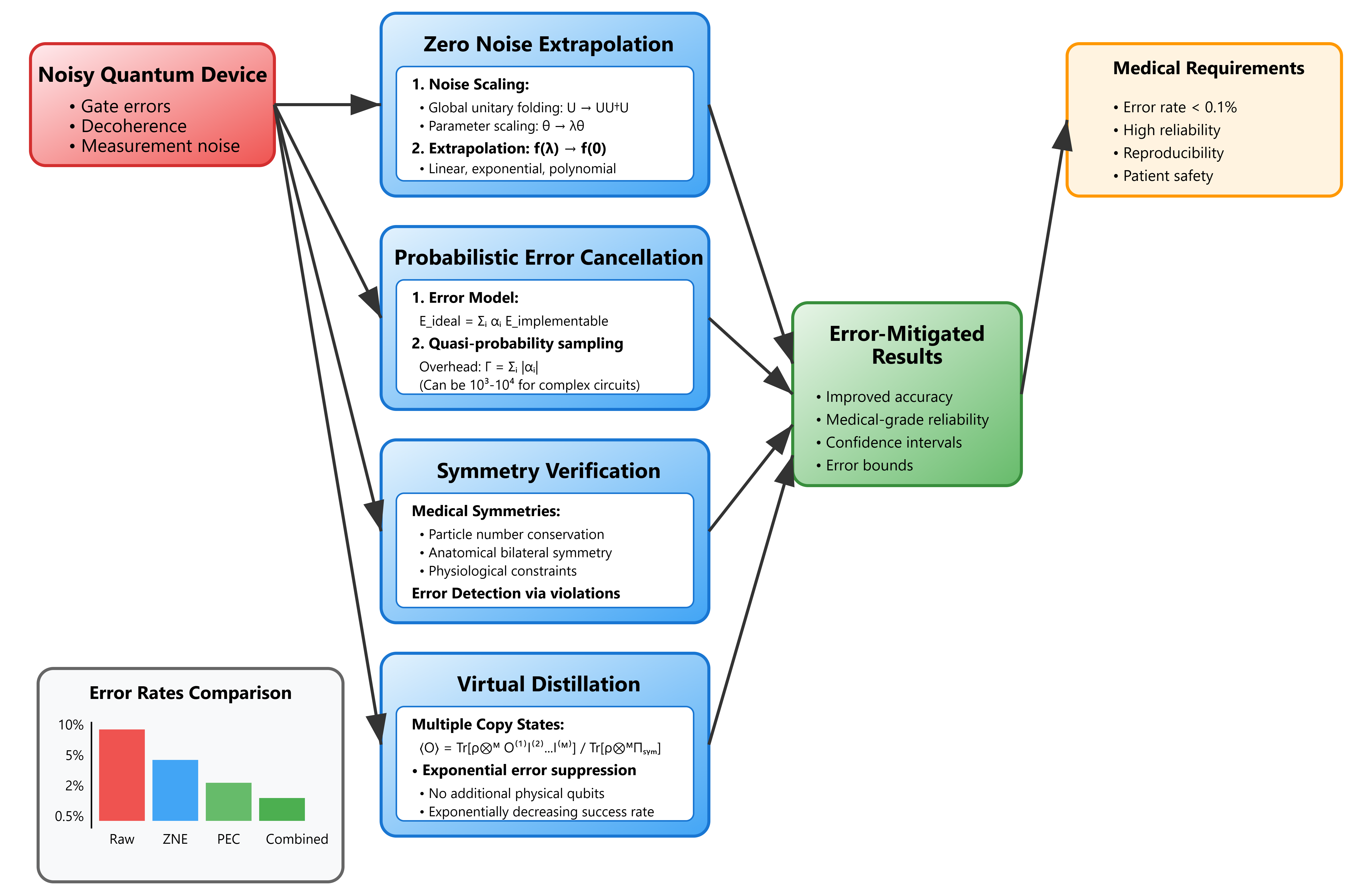}
    \caption{Error Mitigation Workflow for Medical Quantum Computing}
    \label{fig:error_mitigation}
\end{figure}

Medical applications impose stringent reliability requirements that significantly exceed the capabilities of current noisy quantum devices. The probabilistic nature of quantum measurements and the presence of systematic errors from imperfect gates create fundamental challenges for medical applications where reproducible, accurate results are essential for patient safety and clinical decision-making. Fig. \ref{fig:error_mitigation} showcase various error mitigation applications in medical field along with their error rate comparisons, results and requirements. Further, these applications are discussed in detail in subsequent sections.

\subsection{Zero Noise Extrapolation for Medical Applications}

Zero Noise Extrapolation (ZNE) represents one of the most broadly applicable error mitigation techniques for medical quantum computing applications \citep{temme2017error}. The method involves deliberately increasing noise levels in quantum circuits, measuring how computational results degrade with increased noise, and extrapolating back to estimate zero-noise results.

The extrapolation process models the relationship between noise level and computational results through fitting functions that can be extrapolated to the zero-noise limit:
\begin{equation}
\langle O \rangle_{\text{ideal}} = \lim_{\lambda \rightarrow 0} f(\lambda, \{\langle O \rangle_{\lambda_i}\})
\end{equation}
where $\lambda_i$ represent different noise scaling factors and $f$ is the chosen extrapolation function.

To implement ZNE, several noise-scaling techniques have been developed:
\begin{itemize}
    \item \textbf{Global unitary folding:} This method increases circuit noise by replacing each gate $U$ with $UU^\dagger U$, which leaves the logical operation unchanged but introduces additional hardware noise.
    \item \textbf{Local unitary folding:} A more targeted approach, this technique applies the folding selectively to specific gates or subcircuits, allowing finer control over where and how much noise is scaled.
    \item \textbf{Parameter scaling:} For gates that are parameterized (like rotation gates), their parameters can be scaled by a factor $\lambda > 1$, effectively increasing the circuit's sensitivity to noise in a continuous fashion.
    \item \textbf{Digital ZNE:} Instead of modifying circuits directly, this technique uses probabilistic gate substitutions to simulate different noise levels, enabling noise scaling through statistical averaging.
\end{itemize}

Choosing the right extrapolation function is critical. Linear models are simple but might miss nonlinear noise effects. Exponential models better capture real-world decoherence but may be harder to fit accurately. Polynomial models are more flexible but can behave unpredictably if the polynomial degree is too high.

For medical applications, ZNE provides several important advantages. The method is model-agnostic and can be applied to any quantum algorithm without detailed knowledge of the underlying noise processes. ZNE tends to be conservative, typically underestimating rather than overestimating quantum performance, which aligns with medical safety requirements. The extrapolation uncertainty provides confidence intervals that can be interpreted in medical decision-making contexts.

However, ZNE also has significant limitations for medical applications. The measurement overhead scales multiplicatively with the number of noise scaling factors, making real-time medical applications impractical. Extrapolation errors can become large when the signal-to-noise ratio is low, which occurs frequently for complex medical quantum algorithms. The method only corrects for certain types of coherent errors and does not address measurement errors or correlated noise processes that may be particularly problematic for medical applications.

\subsection{Probabilistic Error Cancellation}

Probabilistic Error Cancellation (PEC) offers the theoretical possibility of exact error correction at the cost of dramatically increased sampling overhead \citep{endo2018practical}. The method decomposes noisy quantum operations into linear combinations of implementable operations, enabling exact cancellation of systematic errors through statistical sampling.

The PEC approach expresses ideal operations as linear combinations of noisy operations available on quantum hardware:
\begin{equation}
E_{\text{ideal}} = \sum_i \alpha_i E^{(i)}_{\text{implementable}}
\end{equation}
where the coefficients $\alpha_i$ can be negative, requiring quasi-probability sampling that dramatically increases measurement requirements by a factor of $\Gamma = \sum_i |\alpha_i|$.

For medical applications, the sampling overhead often becomes prohibitive. The overhead factor $\Gamma$ can reach thousands or tens of thousands for complex quantum circuits, making real-time medical applications completely impractical. However, certain medical applications with relaxed timing constraints may find the trade-off between perfect error correction and increased computation time acceptable, particularly for offline analysis or research applications.

Recent studies have looked into ways to lower the cost of PEC by applying it only to the parts of a quantum circuit that are most affected by errors. This selective use helps cut down the number of times the circuit needs to be run while still protecting the most important quantum operations. In medical applications, this means figuring out which parts of the circuit are most important for keeping medical data accurate, which becomes a key step in making the overall process more efficient.

\subsection{Symmetry Verification for Medical Data}

Many forms of medical and biological data inherently possess natural symmetries, which can be leveraged for detecting and mitigating errors in quantum computations. These symmetries, grounded in physical and biological principles, such as conservation laws in molecular systems, anatomical symmetries in medical imaging, and physiological constraints in patient data offers a powerful, domain-informed approach to validating quantum results.

Symmetry verification involves designing quantum circuits and algorithms that preserve known symmetries, and then monitoring for violations that indicate possible computational errors. For example, in quantum simulations of molecular systems, particle number conservation acts as a natural symmetry. Any deviation from the expected particle number during computation may suggest hardware-induced errors. Similarly, in medical imaging, spatial symmetries such as left-right or radial symmetry in organs can be used to perform consistency checks between mirrored or corresponding image regions.

To be effective, symmetry verification must be incorporated at the circuit design stage, not just in post-processing. This means using symmetry-preserving ansätze in variational algorithms, ensuring that the quantum evolution respects the expected symmetries throughout the computation. Doing so not only helps in detecting errors but can also reduce the complexity of the algorithm by shrinking the parameter space, since only symmetry-compliant solutions are explored.

For medical applications, symmetry verification offers several key advantages:
\begin{itemize}
    \item It provides a noise-agnostic error detection mechanism based on fundamental properties of the data, rather than relying on detailed noise models.
    \item Symmetry violations act as strong indicators of faults, enabling real-time adaptive correction or flagging of unreliable results.
    \item The approach is highly compatible with other error mitigation strategies like zero noise extrapolation or readout correction, offering layered protection for high-stakes medical computations.
\end{itemize}

By embedding knowledge of biological or anatomical structure into the quantum algorithm design, symmetry verification ensures that the computational outcomes remain trustworthy and aligned with physical reality, an essential requirement for quantum applications in healthcare and life sciences.

\subsection{Virtual Distillation}

Virtual distillation \citep{koczor2021exponential} offers a powerful way to reduce quantum errors exponentially without needing extra physical qubits, which makes it especially useful in medical applications with limited qubit availability. The technique works by generating multiple identical copies of a quantum state within the computational framework and then applying projections onto symmetric subspaces, effectively filtering out noise and enhancing the fidelity of the final state.

The virtual distillation protocol measures observables on multiple copies of the quantum state:
\begin{equation}
\langle O \rangle_{\text{mitigated}} = \frac{\text{Tr}[\rho^{\otimes M} O^{(1)} \mathbb{I}^{(2)} \cdots \mathbb{I}^{(M)}]}{\text{Tr}[\rho^{\otimes M} \Pi_{\text{sym}}]}
\end{equation}
where $\Pi_{\text{sym}}$ projects onto the symmetric subspace of the $M$-copy system.

The exponential error suppression comes at the cost of exponentially decreasing success probability, creating a trade-off between error reduction and measurement overhead. For medical applications, this trade-off may be acceptable for critical computations where accuracy is paramount and computational time is less constrained.

\section{Molecular Medicine Applications}

Quantum simulation of molecular systems provides some of the most compelling applications for medical quantum computing, leveraging the natural quantum mechanical nature of molecular interactions to achieve computational advantages over classical approaches.

\subsection{Electronic Structure Calculations}

Electronic structure calculations form the foundation of computational drug discovery and molecular medicine applications. The Born-Oppenheimer Hamiltonian describes the electronic structure of molecular systems under the approximation that nuclear motion can be separated from electronic motion:
\begin{equation}
H = -\sum_i \frac{\nabla_i^2}{2} - \sum_{i,A} \frac{Z_A}{|\mathbf{r}_i - \mathbf{R}_A|} + \sum_{i<j} \frac{1}{|\mathbf{r}_i - \mathbf{r}_j|} + \sum_{A<B} \frac{Z_A Z_B}{|\mathbf{R}_A - \mathbf{R}_B|}
\end{equation}
where the terms represent kinetic energy, nuclear-electron attraction, electron-electron repulsion, and nuclear-nuclear repulsion respectively.

The computational complexity of solving the electronic Schrödinger equation scales exponentially with the number of electrons, creating fundamental limitations for classical approaches to large molecular systems. Quantum computers offer the potential to simulate molecular systems with polynomial overhead, though current devices remain limited to small molecular systems due to qubit and coherence constraints.

Active space approximations provide a practical approach to reducing computational complexity by focusing quantum simulations on chemically relevant molecular orbitals. The Complete Active Space (CAS) approach selects $n$ electrons in $m$ molecular orbitals, reducing the full configuration interaction problem to a more manageable size. The dimension of the CAS problem scales as $\binom{m}{n}^2$, which can be efficiently represented on quantum computers for moderate active space sizes.

Recent work has demonstrated quantum simulations of small molecules relevant to drug discovery, including hydrogen molecules, lithium hydride, and beryllium hydride. While these demonstrations involve only a few qubits, they establish the fundamental principles needed for scaling to larger pharmaceutical molecules. The challenge lies in extending these methods to the much larger molecular systems typically involved in drug discovery applications.

\subsection{Drug-Target Interaction Modeling}

Drug-target binding involves complex molecular interactions that present significant challenges for classical simulation methods but may be amenable to quantum simulation approaches. The binding affinity between drug molecules and their protein targets determines therapeutic efficacy and is governed by the free energy change upon binding:
\begin{equation}
\Delta G_{\text{binding}} = G_{\text{complex}} - G_{\text{drug}} - G_{\text{target}}
\end{equation}

Key molecular interactions that contribute to drug-target binding include van der Waals forces arising from quantum mechanical fluctuations in electron density, hydrogen bonding creating directional interactions that affect binding specificity, electrostatic interactions between charged regions of drugs and proteins, charge transfer effects producing electronic coupling between drug and protein, and conformational entropy changes reflecting molecular flexibility and induced fit effects.

Quantum simulations can potentially provide more accurate modeling of these interactions compared to classical force field approaches, particularly for systems where quantum effects such as charge transfer and electronic delocalization play important roles. However, the size of drug-protein complexes typically involves thousands of atoms, far exceeding the capabilities of current quantum devices.

Hybrid approaches that combine quantum simulation of active site regions with classical modeling of protein environments may provide practical pathways for applying quantum methods to drug discovery. These approaches focus quantum computational resources on the most chemically important regions where quantum effects are expected to be most significant.

\subsection{Enzymatic Reaction Mechanisms}

Enzymatic reactions represent another important application area where quantum simulation may provide advantages over classical approaches. Enzymes catalyze chemical reactions through complex mechanisms that often involve quantum effects such as tunneling and electronic reorganization that are difficult to model classically.

The general mechanism of enzymatic catalysis involves substrate binding, transition state stabilization, and product release:
\begin{equation}
\text{Substrate} + \text{Enzyme} \rightleftharpoons \text{ES Complex} \rightarrow \text{Product} + \text{Enzyme}
\end{equation}

Cytochrome P450 enzymes are especially significant targets for quantum simulation because of their vital role in drug metabolism. These enzymes include heme cofactors with intricate electronic structures, often involving transition metals, which make them difficult to simulate accurately using classical approaches.

By leveraging quantum simulations, researchers can gain deeper insights into enzymatic reaction mechanisms, substrate interactions, and metabolic pathways—all crucial for advancing drug discovery and tailoring treatments in personalized medicine. Moreover, understanding how genetic differences in these enzymes influence drug metabolism may lead to more accurate dosing strategies and a reduction in adverse drug reactions.

\section{Medical Imaging Applications}

Medical imaging produces large volumes of high-dimensional data with complex spatial and temporal correlations. These structural patterns creates opportunities for quantum computational approaches for enhanced processing capabilities. The subsequent sections examine quantum-enhanced image reconstruction, classification, and segmentation in medical imaging. 

\subsection{Quantum-Enhanced Image Reconstruction}

Medical image reconstruction is a key step in techniques like MRI and CT, where the final image must be computed from indirect measurements. This process often involves solving large systems of equations of the form $A\mathbf{x} = \mathbf{b}$, where $A$ represents the imaging system, $b$ is the measured data, and $x$ is the image to be reconstructed. Classical methods for solving such systems can be time-consuming, especially as the size of the data increases.

Quantum linear system algorithms (QLSAs) \citep{harrow2009quantum} have been proposed as a faster alternative. These algorithms can, in theory, solve large systems with a runtime that grows only logarithmically with the size of the data, offering a potential speedup compared to classical solvers. However, this advantage comes with important limitations: the system matrix $A$ must meet strict mathematical conditions, and it is often difficult to extract useful information from the quantum solution due to readout challenges. These issues can make it difficult to apply QLSAs directly to real-world medical images.

In many cases, medical images are sparse. This means that most pixel values are zero or near zero in some transform space. This makes compressed sensing techniques a better fit. Compressed sensing tries to reconstruct the image by finding the sparsest solution that still matches the measured data. Quantum algorithms designed for sparse optimization problems can support this process, potentially offering practical benefits for reconstructing high-quality medical images more efficiently.

\subsection{Quantum Image Segmentation}

Medical image segmentation refers to the process of dividing an image into distinct regions that correspond to different tissues, organs, or abnormalities. This task is essential for accurate medical image analysis, diagnosis, and treatment planning. However, the segmentation of complex medical images, which often involve subtle boundaries and overlapping tissues, presents significant computational challenges.

Quantum computing introduces new possibilities by leveraging quantum parallelism to process multiple segmentation pathways simultaneously. One example is the quantum watershed algorithm, which encodes all possible flooding scenarios in a quantum superposition:
\begin{equation}
|\psi\rangle = \frac{1}{\sqrt{N}} \sum_{i=1}^N |i, \text{watershed}(i)\rangle
\end{equation}
Here, each computational basis state corresponds to a disticnt starting point for the watershed process, allowing the algorithm to explore numerous segmentation outcomes in parallel.

Despite its theoretical appeal, the practical realization of quantum image segmentation faces several limitations. The overhead associated with encoding classical image data into quantum states can outweigh the potential computational benefits. Furthermore, extracting segmentation results from quantum states requires measurement operations that may limit the speedup offered by quantum computation.

To address these limitations, hybrid quantum-classical approaches are emerging as a more feasible solution. In such frameworks, quantum algorithms are applied to specific bottlenecks within classical segmentation pipelines, such as optimization or combinatorial search tasks. These hybrid strategies aim to combine the strengths of both paradigms for improved efficiency in medical image segmentation.

\subsection{Medical Image Classification}

Quantum methods for medical image classification are an emerging area of study, with ongoing efforts to evaluate their advantages over classical techniques. Quantum Convolutional Neural Networks (QCNNs) \citep{cong2019quantum} are amongst the prominent architectures being explored. These networks adapt the structure of classical CNNs into a quantum framework using a sequence of quantum convolution and pooling operations:
\begin{equation}
|\psi_{\text{out}}\rangle = \prod_{\text{layers}} U_{\text{pooling}} U_{\text{conv}}(\theta) |\psi_{\text{in}}\rangle
\end{equation}
In principle, such quantum circuits can access exponentially large feature spaces, which may enable the discovery of complex patterns not easily captured by classical models. However, the utility of these high-dimensional quantum features in practical medical image classification remains uncertain, particularly given the current hardware limitations. Existing implementations are restricted to low-resolution images and basic classification problems.

An alternative and potentially more practical direction involves quantum kernel methods. These approaches leverage quantum circuits to compute inner products (kernels) between image data points in high-dimensional Hilbert spaces. Such quantum kernels can capture intricate correlations and can be used in conjunction with classical ML algorithms, offering a hybrid strategy that benefits from quantum-enhanced feature representations.

The practical impact of quantum image classification will likely depend on identifying specific medical imaging problems where quantum processing offers measurable improvements over classical deep learning. This will require a detailed understanding of the structural characteristics of medical images and the diagnostic relevance of various feature types.

\section{Implementation Challenges and Future Directions}

Quantum computing holds transformative potential for biomedical data processing and medical diagnosis. However, its adoption in real-world clinical environments faces several challenges. This section categorizes these limitations into four areas: hardware constraints, system integration, regulatory concerns, and directions for future research.

\subsection{Current Hardware Limitations}

Current quantum processors are restricted by the capabilities of NISQ devices. These limitations hinder the ability to execute complex quantum circuits required for high-dimensional biomedical data analysis. Table \ref{tab:hardware} outlines major limitations in today’s quantum hardware platforms.

\begin{table}[h!]
\centering
\scriptsize
\caption{Hardware limitations of current quantum systems}
\begin{tabular}{p{2.5cm}p{13cm}}
\toprule
\textbf{Limitation} & \textbf{Description} \\
\midrule
Limited Qubit Count & Most devices support between 50 and 1000 qubits, constraining their use for large-scale medical datasets. \\
\midrule
Qubit Connectivity & Most hardware allows only nearest-neighbor interactions, requiring additional \text{SWAP} gates that increase circuit depth. \\
\midrule
Short Coherence Time & Qubits maintain quantum states only for microseconds to milliseconds, limiting the execution of complex algorithms. \\
\midrule
Gate Errors & Quantum gates typically have error rates around 0.1–1\%, degrading overall fidelity. \\
\midrule
Limited Circuit Depth & Accumulation of noise restricts the number of quantum operations that can be performed before decoherence degrades the quantum state. \\
\bottomrule
\end{tabular}
\label{tab:hardware}
\end{table}

\subsection{Integration with Classical Medical Systems}

Integrating quantum computing into classical healthcare infrastructure presents significant technical and operational challenges. For quantum medical applications to be clinically viable, they must interoperate seamlessly with essential systems such as EHR, medical imaging platforms, laboratory information systems (LIS), and clinical decision support systems (CDSS).

A key challenge lies in reconciling the probabilistic nature of quantum computation with the deterministic expectations of clinical workflows. Medical decision-making demands consistent and reproducible outputs. Therefore, quantum algorithms must incorporate post-processing techniques, error mitigation strategies, and statistical validation to produce clinically acceptable results.

Time-sensitive applications such as emergency care, intensive monitoring, and intraoperative support require real-time or low-latency computation. However, current quantum hardware is constrained by limited coherence times and significant measurement overhead, making it difficult to meet these demands. Hybrid quantum-classical approaches tailored for real-time responsiveness are necessary to address this issue.

Data compatibility is another concern. Clinical systems rely on standardized data formats such as DICOM for imaging, HL7 for clinical records, and FASTQ/VCF for genomic data. Quantum algorithms must be integrated with data preprocessing pipelines capable of translating these formats into quantum-compatible representations without disrupting existing workflows.

To address these challenges, future work should focus on:
\begin{itemize}
    \item Developing robust interfaces between quantum algorithms and classical medical systems (e.g., EHR, LIS, CDSS).
    \item Ensuring consistency in clinical outputs through statistical validation and hybrid processing.
    \item Designing low-latency quantum-classical architectures for time-critical healthcare applications.
    \item Creating efficient pipelines to convert standardized medical data into quantum-readable formats.
\end{itemize}

\subsection{Regulatory and Clinical Validation}

As quantum computing applications in medicine advance beyond theoretical exploration and prototype demonstrations, their translation into clinical practice must align with established regulatory standards. Regulatory compliance ensures that quantum medical algorithms uphold the safety, efficacy, and accountability standards essential to patient care. However, given the novel computational paradigm introduced by quantum systems, significant gaps exist in current regulatory frameworks.

In the United States, the Food and Drug Administration (FDA) regulates Software as a Medical Device (SaMD), which may encompass quantum medical software in the future. While the FDA has published guidelines for AI/ML-based medical software, explicit pathways for quantum algorithms are still under development. Table \ref{tab:regulatory_challenges} outlines the regulatory challenges and required adaptations.

Clinical validation of quantum algorithms extends beyond functional correctness. It must demonstrate measurable clinical benefit through well-designed clinical studies. This includes phases such as technical benchmarking, safety profiling, and randomized clinical trials (RCTs) to compare quantum-assisted interventions against standard-of-care methods. The intrinsic probabilistic outputs and hardware noise sensitivity of quantum systems pose substantial complications for traditional clinical validation processes.

Moreover, current software validation strategies, which assume deterministic behavior, are not readily applicable to quantum software. This necessitates the development of quantum-aware validation frameworks that can rigorously test outcome reliability and statistical significance under quantum uncertainty.

To meet the clinical standards required for deployment in hospitals and diagnostic centers, quality assurance procedures must incorporate confidence interval estimation, reproducibility checks, and robustness analysis against quantum noise. These measures will be essential to demonstrate that quantum outputs are sufficiently stable and interpretable for use in high-stakes clinical environments.

\begin{table}[h!]
\centering
\scriptsize
\caption{Regulatory Challenges for Quantum Medical Algorithms}
\label{tab:regulatory_challenges}
\begin{tabular}{p{3.5cm}p{12.5cm}}
\toprule
\textbf{Challenge} & \textbf{Description and Requirements} \\
\midrule
Lack of Quantum-Specific Guidance & Regulatory bodies such as the FDA lack frameworks specific to quantum medical software; adaptation of SaMD policies is needed. \\
\midrule
Probabilistic Outputs & Quantum algorithms produce stochastic results, requiring validation strategies that account for uncertainty and noise. \\
\midrule
Hardware Dependence & Validation must consider platform-specific noise models and hardware-specific performance, impacting reproducibility. \\
\midrule
Algorithm Interpretability & Regulatory approval demands explainability, which is complicated by non-intuitive quantum mechanisms. \\
\midrule
Validation Protocols & Traditional software QA protocols do not translate directly to quantum workflows; new metrics and protocols are required. \\
\bottomrule
\end{tabular}
\end{table}

Future regulatory pathways must evolve alongside technological advances in quantum computing. Establishing industry-wide standards, quantum reliability benchmarks, and medical-grade certification processes will be key to integrating quantum technologies into real-world clinical settings.

\subsection{Research Priorities and Future Developments}

To advance the integration of quantum computing into medical applications, several research directions must be prioritized:

\begin{itemize}
    \item \textbf{Domain-Specific Algorithm Design:} 
    \begin{itemize}
        \item Develop quantum algorithms tailored to the structure and characteristics of medical data.
        \item Incorporate domain expertise from medical physics, biological systems, and clinical practice.
        \item Move beyond generic quantum routines to problem-specific formulations.
    \end{itemize}

    \item \textbf{Quantum Error Correction for Medical Reliability:}
    \begin{itemize}
        \item Design error correction codes that ensure reliability levels suitable for clinical applications.
        \item Prioritize correction of error types that significantly impact medical diagnosis or treatment.
        \item Adapt existing protocols to account for real-time reliability requirements in healthcare settings.
    \end{itemize}

    \item \textbf{Application-Specific Hardware Development:}
    \begin{itemize}
        \item Build quantum hardware optimized for medical tasks rather than general-purpose use.
        \item Design processors with connectivity, gate sets, and error correction schemes that align with medical imaging and diagnostic workflows.
    \end{itemize}

    \item \textbf{Hybrid Quantum-Classical Algorithm Development:}
    \begin{itemize}
        \item Create algorithms that combine classical and quantum computation for practical utility on near-term quantum devices.
        \item Identify computational sub-tasks in medical pipelines that can benefit most from quantum acceleration.
        \item Ensure modular integration to allow seamless adoption in clinical workflows.
    \end{itemize}

    \item \textbf{Validation and Verification Frameworks:}
    \begin{itemize}
        \item Develop robust frameworks for testing and validating quantum algorithms in medical contexts.
        \item Ensure compliance with clinical standards of accuracy, safety, and interpretability.
        \item Introduce standardized benchmark datasets, performance metrics, and reproducibility protocols.
    \end{itemize}
\end{itemize}

\section{Conclusion}



Quantum state preparation for medical data is a growing field that combines quantum computing and healthcare technology. While researchers have strong theoretical foundations and advanced mathematical methods to encode medical data into quantum states, putting these ideas into practice to improve patient care remains challenging.

Medical data has unique properties, such as spatial patterns in images, layered structures in biological systems, and time based trends in health signals. These features make the data well suited for quantum encoding techniques like tensor networks, variational algorithms, and quantum machine learning. Since these properties align with problems where quantum computers may have an advantage, there is optimism about future medical applications of quantum technology.

However, current hardware limitations severely constrain the scope of practical applications. The limited number of qubits, high error rates, and short coherence times of contemporary quantum devices restrict implementations to small-scale problems that do not capture the full complexity of real medical systems. Error mitigation techniques provide temporary solutions for improving computational accuracy, but fault-tolerant quantum computing will likely be required for many medically relevant applications.

The timeline for meaningful clinical impact extends over decades and requires continued advances in multiple complementary areas. Quantum hardware must achieve substantial improvements in qubit count, gate fidelity, and coherence time to enable practical medical applications. Algorithm development must focus on problem-specific approaches that efficiently exploit medical data structure while working within hardware constraints. Integration frameworks must be developed to connect quantum algorithms with classical medical systems and existing healthcare infrastructure.

Near-term research should emphasize proof-of-concept demonstrations that establish the feasibility of quantum approaches for specific medical problems, validation methodology development that ensures clinical safety and reliability, and hybrid classical-quantum approaches that provide incremental improvements while quantum hardware continues maturing. Success will be measured not by abstract quantum advantage demonstrations, but by genuine improvements in diagnostic accuracy, therapeutic effectiveness, and patient outcomes.

The ultimate goal remains clear: leveraging quantum computational advantages to improve human health through better diagnostics, more effective treatments, and accelerated medical research. Achieving this goal requires sustained collaboration between quantum computing researchers, medical professionals, and healthcare technology developers, always maintaining patient benefit as the primary objective. While the challenges are substantial, the potential for transformative healthcare improvements through quantum computing justifies continued research investment and development effort.

\bibliographystyle{unsrtnat}
\bibliography{ref}

\end{document}